\documentclass[preprint,preprintnumbers,amsmath,amssymb]{iopart}
\usepackage{graphicx}
\usepackage{fullpage}

\newcommand{\pc}{\%~}

\begin{document}

\title[Gyrokinetic modelling of impurity transport including rotation in AUG]{Validation of gyrokinetic modelling of light impurity transport including rotation in ASDEX Upgrade}

\author{F.J.~Casson, R.M.~McDermott, C.~Angioni, Y.~Camenen$^1$, R.~Dux, E.~Fable, R.~Fischer, B.~Geiger, P.~Manas$^1$, L.~Menchero$^{(2)}$, G.~Tardini, and ASDEX Upgrade team}
\address{Max-Planck-Institut f\"{u}r Plasmaphysik, IPP-EURATOM Association, D-85748 Garching bei M\"{u}nchen, Germany}
\address{$1$ Aix-Marseille Universit\'{e}, CNRS, PIIM UMR 7345, 13397 Marseille, France}
\address{$(2)$ ADAS, Department of Physics, University of Strathclyde, 107 Rottenrow East, Glasgow G4 0NG, United Kingdom}



\begin{abstract}

Upgraded spectroscopic hardware and an improved impurity concentration calculation
allow accurate determination of boron density in the ASDEX Upgrade tokamak.  
A database of boron measurements is compared to 
quasilinear and nonlinear gyrokinetic simulations including Coriolis and centrifugal rotational effects over a range of H-mode plasma regimes.  
The peaking of the measured boron profiles shows a strong anti-correlation 
with the plasma rotation gradient, via a relationship explained and reproduced by the theory.
It is demonstrated that the rotodiffusive impurity flux driven by the rotation gradient
is required for the modelling to reproduce the hollow boron
profiles at higher rotation gradients.  The nonlinear simulations validate the quasilinear approach, 
and, with the addition of perpendicular flow shear, 
demonstrate that each symmetry breaking mechanism that causes
momentum transport also couples to rotodiffusion.  At lower rotation gradients, the parallel compressive
convection is required to match the most peaked boron profiles.
The sensitivities of both datasets to possible errors 
is investigated, and quantitative agreement is found within the estimated 
uncertainties. 
The approach used can be considered a template for mitigating uncertainty 
in quantitative comparisons between simulation and experiment.

\end{abstract}


\maketitle

\section{Introduction}

To achieve optimum fusion performance, future reactors will need to control 
the build up of impurities in the plasma core.  Light impurities dilute the fusion fuel, while
heavier impurities determine radiation losses and impact plasma control.
The design and operation of future reactors will therefore benefit greatly from predictive capability
for modelling impurity transport under different plasma conditions.

Core impurity transport is observed to be anomalous in both H modes and L modes
\cite{guirlet_parametric_2006,valisa_metal_2011,howard_quantitative_2012}, 
though neoclassical transport can also contribute,
 particularly for heavier impurities and at higher collisionalities.
In the edge, transport of impurities has been observed to follow 
the neoclassical predictions \cite{putterich_elm_2011}. 
The tools for modelling both neoclassical and turbulent impurity transport have reached the stage
of development at which it is possible to make comparisons with experiment.  Quantitative comparisons 
across a range of plasma conditions are necessary to validate the appropriateness of these tools before they can
be applied with confidence to make predictions for future devices \cite{greenwald_verification_2010}.

Turbulent transport is well known to be stiff with respect to the drive, making quantitative comparison of fluxes challenging in all 
transport channels.  This difficulty can be avoided by the examination of steady state phases 
in which dimensionless ratios of transport coefficients are used to predict null flux density gradients
which can be compared directly with the measurements.  
For impurities, one obstacle to such efforts is the difficulty of making accurate measurements of impurity density, 
but recent studies are starting to demonstrate promising qualitative \cite{valisa_metal_2011,nordman_fluid_2011,angioni_gyrokinetic_2011} and quantitative agreement \cite{howard_quantitative_2012,howard_measurement_2012} between between impurity measurements and gyrokinetic turbulence simulations.

The aim of this work is to quantitatively compare high quality measurements 
to physically comprehensive gyrokinetic and neoclassical modelling including all the effects of rotation.  
The gyrokinetic modelling is performed with the {\sc gkw} code \cite{peeters_nonlinear_2009} including the rotodiffusive contribution \cite{camenen_impact_2009} (the off-diagonal coupling between rotation gradient and particle transport channels), Coriolis \cite{peeters_toroidal_2007,peeters_influence_2009}
and centrifugal effects \cite{peeters_influence_2009,casson_gyrokinetic_2010}. The neoclassical transport is computed
with the {\sc neo} code \cite{belli_kinetic_2008,belli_eulerian_2009,belli_full_2012} (which also contains the effects of rotation), but is found to be negligible for the database considered.  
We exploit the recently upgraded charge
exchange (CX) recombination spectroscopy diagnostic on the ASDEX Upgrade tokamak (AUG) combined with recent improvements
in the density interpretation \cite{dux_impurity_2012} to build a database of boron profiles for H--mode plasmas
under a variety of heating regimes.  This work builds on the work of Ref. \cite{angioni_gyrokinetic_2011} 
by adding a more diverse database covering a larger parameter range, with more accurate measurements.  
The present modelling furthermore includes centrifugal effects (found to be only a minor correction in this dataset) and a
detailed investigation of the sensitivities of the results to enable quantitative comparison allowing identification of distinct physical mechanisms.
The sensitivity study is conducted for quasilinear local simulations of density peaking for the entire database, and is complementary to the approach used in Ref. \cite{howard_quantitative_2012}, in which sensitivity studies were performed for a single case to match both convective and diffusive contributions determined independently from a dedicated perturbative experiment.
      
Together, these improvements allow different off-diagonal impurity transport mechanisms to be unambiguously distinguished in experiment for the first time.
We show that, within the uncertainties, the primary correlation in the database, between plasma rotation gradient and impurity peaking,
is quantitatively reproduced and explained by the modelling only when the rotodiffusive contribution is included.
Nonlinear simulations are presented for a subset of points to validate the quasilinear approach, and to demonstrate that the symmetry breaking effect
of background $E \times B$ shearing increases the rotodiffusive transport.

This paper is structured as follows:  Section II describes the improvements to the CX measurement of 
boron density, Section III presents the experimental database, and the investigation of systematic measurement uncertainties.  Section IV describes the modelling, including sensitivities studies, to enable a robust quantitative comparison.  Section V presents the nonlinear simulations, and conclusions are drawn in Section VI.

\section{Measurement of Boron Density}

The core charge exchange (CX) diagnostic on AUG was substantially upgraded prior to the 2011 campaign
to a 30 channel system which provides much higher temporal resolution and more complete profiles as compared with the previous system
\cite{viezzer_high-resolution_2012,mcdermott_core_2011}. This system is routinely used to measure intrinsic boron (the tungsten wall is regularly boronised) to provide high-resolution measurements of ion temperature and rotation. Impurity densities can also be calculated from the CX spectra, but 
this analysis is more challenging, not only because spectral intensities must be correctly calibrated, but also because the subsequent calculation requires an accurate determination of the neutral population in the plasma including the different energy components of the neutral beam, and the halo population up to at least the first excited state. 
The halo population consists of thermal neutrals originating from CX reactions between plasma deuterium ions and injected beam neutrals.  In addition to the beam neutrals, the halo neutrals can also undergo thermal CX reactions with the impurities in the plasma, and thereby contribute to the total active CX signal (here, ``active'' means photon emission induced by the beam and its halo neutrals, and ``passive'' means photon emission from the plasma edge still present when the beam is off).  Previously, on AUG, the contribution of the halo neutrals to the CX signal was neglected, because halo population was assumed to be small.
Recently, however, a new beam emission spectroscopy (BES) system was implemented on AUG, which revealed a larger halo population \cite{dux_impurity_2012}
than previously expected.  The excited halo population should be taken into account in the CX analysis, because the cross-sections for the first excited state (n=2), are two orders of magnitude larger than the ground state cross-sections for NBI relevant energies (60keV) \cite{guzman_calculation_2010}.  
Therefore, even a small fraction of the halo population in the first excited state can provide a number of active CX photons comparable to the first energy component of the NBI, and should be considered in the CX analysis of boron densities.

\begin{figure}
\centering
\includegraphics[width=85mm]{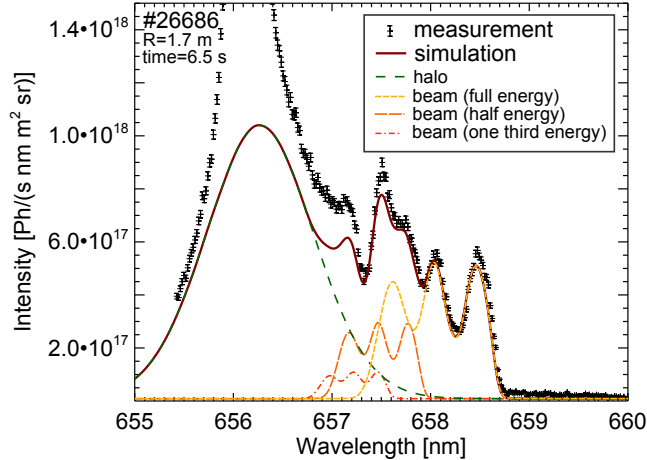}
\caption{Example validation of the {\sc fidasim} simulation of the neutral beam (lines), by comparison with the measured beam emission spectrum (points with error bars). The
total simulated emission spectrum (solid line) closely matches the measured spectrum above 657 nm, indicating that all the beam components and halo are accurately modelled.  At shorter wavelengths, the discrepancy between the simulated and measured spectra is due to passive background emissions from the plasma edge, which are not simulated.}
\label{fig.bes}
\end{figure}

The AUG CX impurity concentration analysis code ({\sc chica}) has therefore been coupled to the Monte-Carlo beam simulation code {\sc (f90)fidasim} \cite{heidbrink_code_2011,geiger_fast-ion_2011,geiger_investigation_2012} which models (on a 3D grid) the neutral beam attenuation, the halo formation and the corresponding spectra including excited states up to n=6.  The coupling to {\sc chica}  provides more accurate modelling of the beam components and allows inclusion of the halo contribution.  The attenuation of the neutral beam from {\sc fidasim} has been compared with both {\sc chica} and {\sc transp} \cite{pankin_tokamak_2004} and found to be in good agreement.
The best evidence for the reliability of {\sc fidasim} is how well it reproduces the experimentally measured BES D-alpha spectra, which include contributions from all the neutral populations present in the plasma, including the halo \cite{geiger_investigation_2012}.  An example of this agreement  is shown in Fig. \ref{fig.bes} for one of the discharges in this work.
In addition to the coupling with {\sc fidasim}, the version of {\sc chica} used in this work also includes ADAS (Atomic Data and Analysis Structure) thermal CX effective emission rates \cite{hoekstra_charge_1998,summers_h._p._adas_2004} for the evaluation of the halo contribution to the CX signal, and has been upgraded with an improved NBI geometry based on BES measurements.  The {\sc fidasim} simulations together with the ADAS data also allow verification that the ground state halo contribution to the CX signal can be safely ignored, as it only contributes $\sim$ 1\pc to the active signal due to the low CX cross-sections at thermal energies.
Overall, these improvements have reduced the calculated boron densities to $\sim 40\%$ of their value in earlier analyses which did not include the halo.  The changes in the logarithmic gradient scale lengths are less dramatic, since as a crude approximation, the halo density is proportional to the beam neutral density.  The values of $R/L_{nB}$ at mid radius are reduced by less than 1.0, which is within the sample error bar given in Ref. \cite{angioni_gyrokinetic_2011}.

\begin{figure}
\centering
\includegraphics[width=120mm]{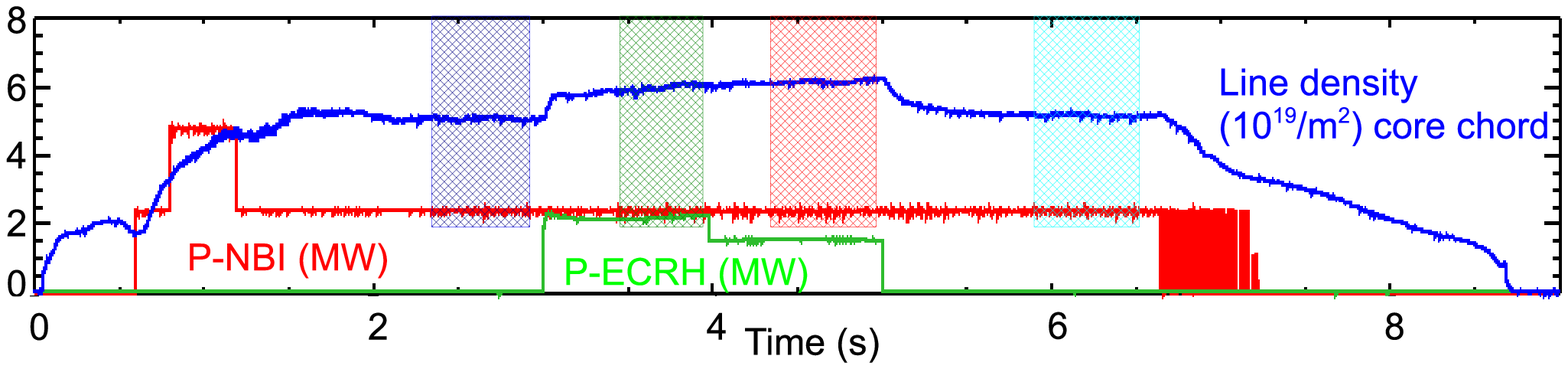}
\includegraphics[width=121mm]{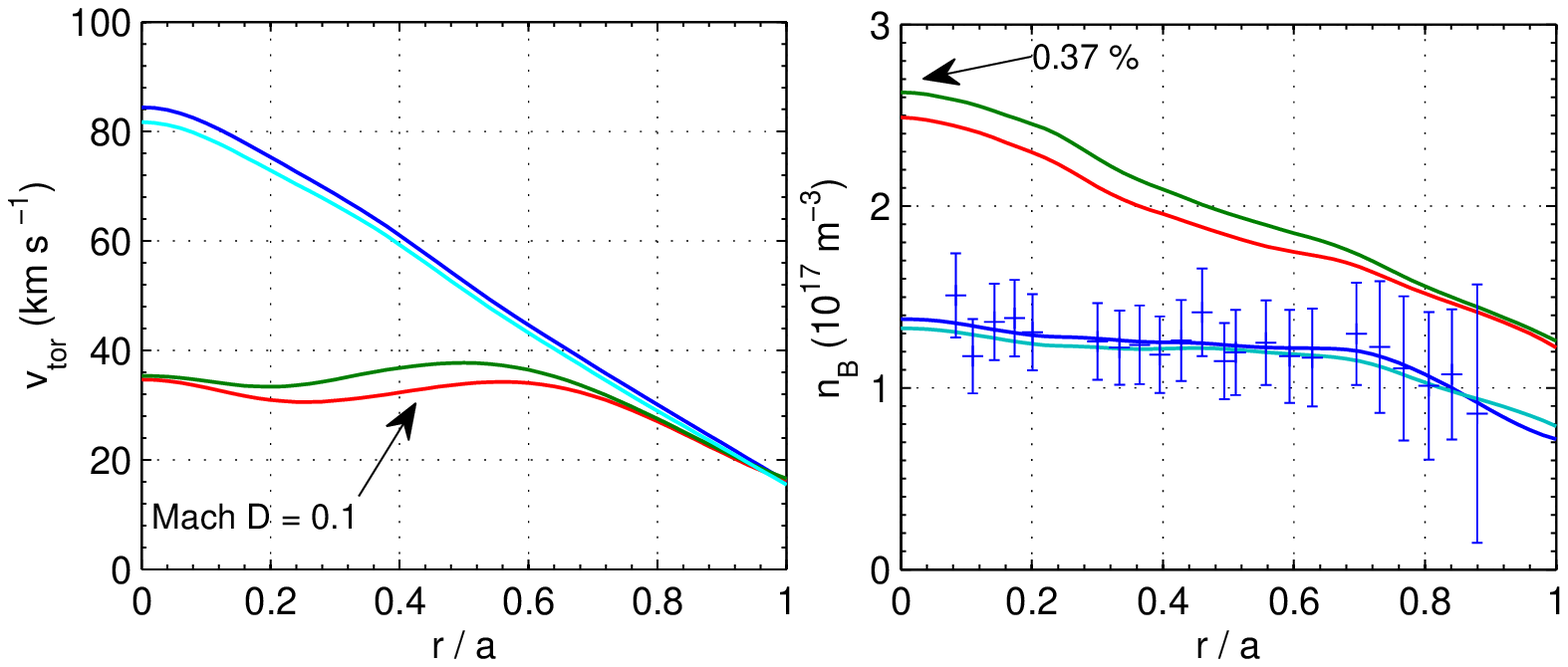}
\caption{Example shot 26686 from the AUG boron database with $I_p=600 kA$.  The 4 profiles of boron rotation and density correspond to the time averages over the colored hatched phases in the timetrace.  Example raw data points with error bars are shown for one of the fits.}
\label{fig.example}
\end{figure}

The phases and profiles in a typical shot are shown in Fig. \ref{fig.example}, which provides an example of a somewhat generic result for AUG:  When electron heating is applied to an NBI heated H-mode, increased density peaking and decreased rotation are observed \cite{mcdermott_effect_2011}. 

\section{Experimental database}

For this work, a database of boron density profiles from H-mode shots from the 2011 campaign on AUG has been constructed.
The database consists of 29 steady phases of at least 0.25 seconds from 10 shots; the energy confinement time for these phases ranges from 0.06 to 0.12 seconds.  All shots have a continuous 2.5 MW NBI heating power from the beam used for the CX measurements, which is predominantly radial (major radius of tangency / major radius of axis = $R_T/R_A = 0.31$).  This heating provides a minimum to which additional NBI and ECRH heating are added (shots with ICRH are excluded). The database covers a range of plasmas with plasma currents $I_p$ from 0.6 to 1.0 MA, NBI powers between 2.5 and 7.5 MW, ECRH powers from 0 to 2.5 MW, core electron densities between 5 and 12 $\cdot 10^{-19} m^{-3}$, Greenwald fraction $n_e/n_{GW}$ = 0.35-0.85, $\beta_N$ = 0.65-2.0, and safety factor $q_{95}$ = 3.9-7.1.  All plasmas are lower single null with toroidal field -2.5T and similar edge elongation (1.6) and average triangularity (0.15).  This database covers a greater range of NBI heating and Mach numbers than Ref. \cite{angioni_gyrokinetic_2011} which also used earlier charge exchange hardware and did not use
the improved analysis for the impurity concentration.

\begin{figure}
\centering
\includegraphics[width=160mm]{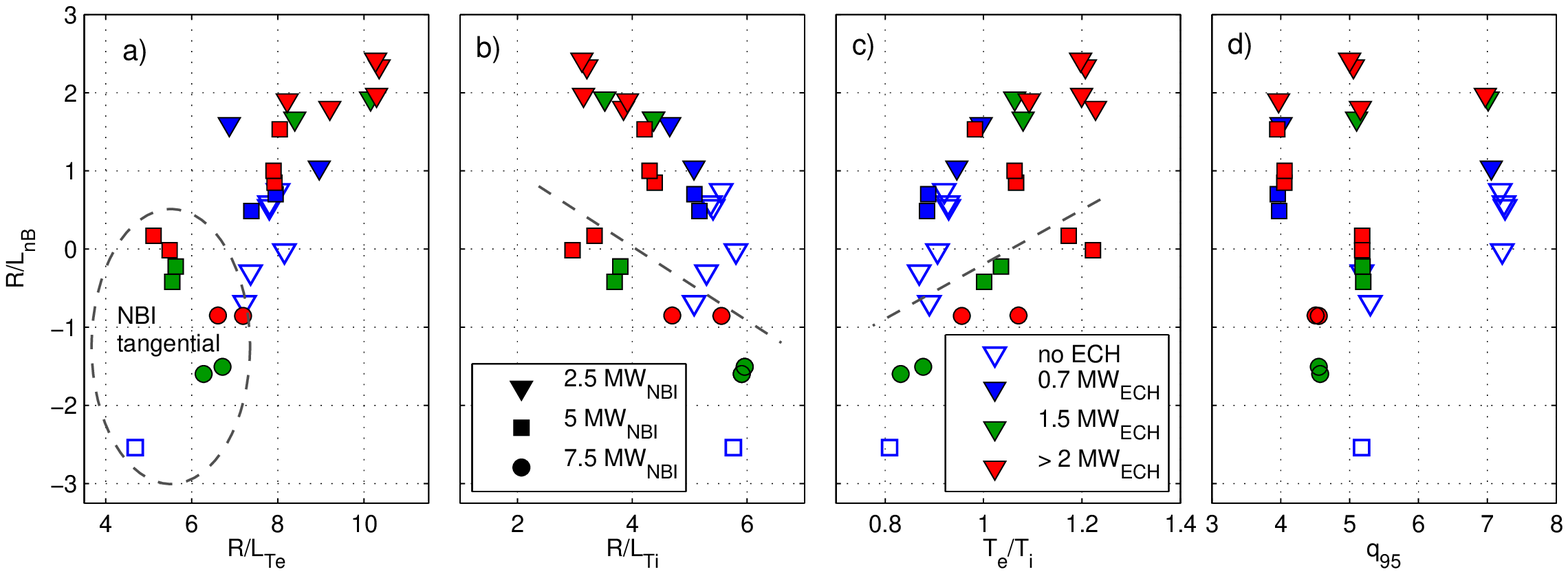}
\includegraphics[width=160mm]{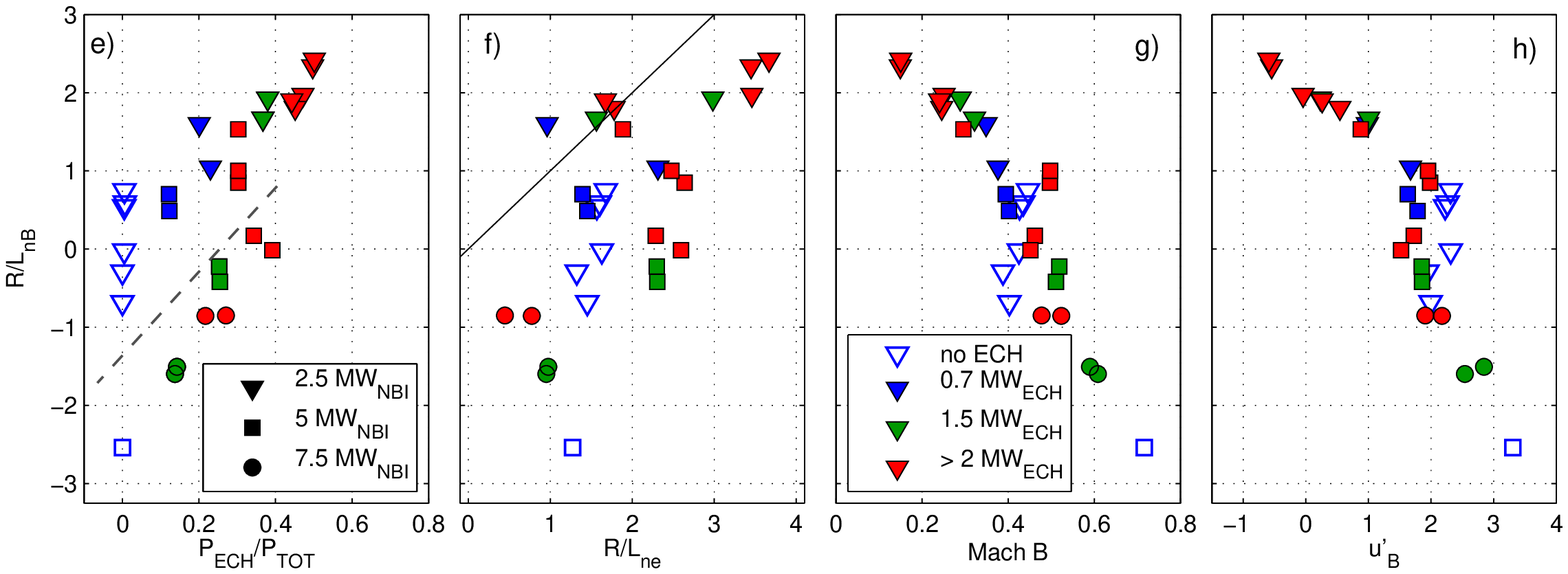}
\caption{Correlations in the AUG experimental H-mode database of boron density measurements at $r/a=0.5$.  The color of the points indicates the ECRH heating power between 0 and 2.5 MW, the shape of the points represents the NBI heating power between 2.5 and 7.5 MW.  For the points inside or below the dashed line, the most tangential NBI beam provides NBI heating in addition to the base 2.5 MW.  The black line in plot (f) is the line of equivalence.} 
\label{fig.scatter}
\end{figure}

A number of strong correlations are immediately visible in the database (Fig. \ref{fig.scatter}).  The most striking is between the Mach number $ u=v_{tor} / v_{th}$ and the logarithmic gradient of the boron density, $R/L_{nB}$ (Fig 3g). The Mach number is defined with the species thermal velocity $v_{th} = \sqrt{2 T / m}$.  A similarly strong correlation can also be seen between $R/L_{nB}$ and the rotation gradient $u^\prime = -({R^2/v_{th_i}}){d \omega / d r}$ (Fig 3h, where $u^\prime_B$ is normalized with the boron thermal velocity).  Here, $\omega$ is the plasma angular rotation frequency.
The similarity between Figs 3g and 3h is due to a high correlation between $u$ and $u^\prime$ in this dataset, which is a consequence of peaked rotation profiles that result from central NBI heating, making the $u$ and $u^\prime$ dependencies indistinguishable at mid radius.  Since temperature gradients are the primary quantity in determining the characteristics of turbulent modes, the correlations between temperature gradients and $R/L_{nB}$ provide a strong signature that turbulence is important in the impurity transport.  Clearly, many of the remaining ``independent variables'' plotted on the x-axes also have correlations between them (e.g. heating powers, temperature gradients and rotation):  A strong correlation between rotation and ion temperature gradient (not shown), is present (since the AUG NBI heating systems always drive a co-current rotation),  but contains two branches due to different injection angles and deposition radii of the NBI beams.  The different geometries of the NBI beams are also responsible for the two branches visible in the plots of $R/L_{nB}$ against $R/L_{Ti}$ and $T_e/T_i$ and $P_{\rm EC}/P_{\rm tot}$ (Figs, 3b, 3c, 3e respectively); the points with more than 2.5 MW NBI in the highlighted branch all include the most tangential beam (which has $R_T/R_A= 0.75$).   Fig. 3d indicates that the safety factor is not a primary dependence, which suggests that neoclassical transport is not dominant.  

The primary correlation, between rotation gradient and impurity peaking is not caused by a single mechanism, but  
is a non-trivial combination of the correlations between heating powers, temperature gradients, and rotation 
(some of which are a result of the beam geometries specific to AUG).  These quantities together determine
the turbulence characteristics (as will be demonstrated by the modelling in the following sections) 
in such a way that the overall result produces the strong correlation shown.

The comparison of the simulations with experiment over a sufficiently large database helps to avoid random errors in the measurements from distorting the global comparison over the database.  Systematic errors, by contrast, are much harder both to estimate and to eliminate, and can certainly affect 
the quantitative comparison between measurement and modelling.  In the Appendix we identify and estimate the largest sources of known uncertainty in the boron density measurement, and detail the uncertainty estimation for the logarithmic gradient.  The sensitivities of the modelling to input uncertainties are investigated in Sec. \ref{sec.modelling}.

\section{Quasilinear Gyrokinetic modelling}

To validate our understanding of the physical mechanisms responsible for the trends in the database, quasilinear simulations were conducted for each point in the database with the flux tube version of the gyrokinetic code {\sc {\sc gkw}} \cite{peeters_nonlinear_2009}.  The boron is modelled in the trace limit in which the micro-instabilities are completely determined by the bulk plasma conditions.  In this limit, the response of impurities to the turbulence is completely linear in the gradients of the impurities, even for nonlinear simulations \cite{angioni_direction_2006,angioni_gyrokinetic_2009,angioni_particle_2009-1}.  The impurity flux can then be exactly decomposed as 
\begin{equation}
\label{eq:particle}
\Gamma_s = n_s D_s \left ({R \over L_{ns}} \biggr|_{R_{LFS}} + C_T {R \over L_{Ts}} + C_u u^\prime_s + C_p \right)
\end{equation}
where the individual terms correspond to the diffusive, thermodiffusive, rotodiffusive, and convective parts, respectively. Here we define the rotodiffusive coefficient $C_u$ using the impurity rotation gradient $u^\prime_s$ normalized with the thermal velocity of the impurity ions.
The rotodiffusive term \cite{camenen_impact_2009} is a result of symmetry breaking mechanisms and is therefore closely connected with the mechanisms of momentum transport \cite{peeters_overview_2011,angioni_off-diagonal_2012}.  The trace impurity transport dimensionless coefficients $C_T, C_u$, and $C_p$ are properties of only the background turbulence, and can be calculated from 4 trace species (all boron in this case) with an (arbitrary) orthogonal set of gradients (explicit formulae are given in Ref. \cite{casson_gyrokinetic_2010}).  Inserting the temperature 
and rotation gradients of the boron measured via charge exchange, we obtain a prediction for the steady state ($\Gamma_B = 0 $) density gradient which can be directly compared to the  independently measured value.  The predicted gradient is a combination of terms of different signs, requiring accurate treatment of symmetry breaking mechanisms and providing a sensitive test of the turbulence theory.

The comparison of dimensionless parameters is particularly suited to quasilinear modelling, since only the shape and not the magnitude of the spectrum has an influence.  The quasilinear results are obtained using a spectrum of 5 modes at $k_\theta \rho_i = [0.2,0.3,0.4,0.5,0.6]$ in which the $k_\theta \rho_i = 0.3$ mode dominates (the choice of spectra is justified Sec. \ref{sec.nonlinear}).  Here, $k_\theta \rho_i = k_\theta \rho_s \sqrt{T_e/T_i}$ where $\rho_s= c_s / \omega_{ci} $ is evaluated using the field at the magnetic axis
[\footnote{In general geometry, the projection of $k_\theta$ onto the binormal coordinate uses the metric tensor \cite{peeters_nonlinear_2009}, not all codes do this in the same way.}].
The spectral shape is fixed the same for all cases and is compared with representative nonlinear simulations in Fig. \ref{fig.nonlinear} (discussed further in Sec. \ref{sec.nonlinear}).  This range of $k_\theta$ was used because wider ranges suffered from the problem of too many linearly stable modes.

All the gyro-kinetic simulations include collisions (pitch angle scattering, with $Z_{\rm eff}= 1 + (1.8e19/n_e)^{0.6}$), full experimental flux surface geometry, and are performed in the frame of reference that rotates with the plasma, keeping both centrifugal and Coriolis forces \cite{peeters_toroidal_2007,peeters_influence_2009,casson_gyrokinetic_2010} and including the drive from the gradient in the plasma rotation \cite{peeters_linear_2005,peeters_erratum:_2012,casson_erratum:_2012}.  The plasmas are modelled as pure plasmas with $n_i=n_e$ and the boron impurity in trace concentration.  While dilution effects may result in a reduction in the turbulence level \cite{howard_quantitative_2012,szepesi_analysis_2013}, they do not have much influence on the steady state density gradient, since it is a ratio of convective to diffusive transport components.  
Electromagnetic fluctuations are neglected and are expected to give only minor impurity profile flattening for these cases \cite{hein_electromagnetic_2010} (which mostly have $\beta_e < 0.5 \%$).  

The influence of neoclassical transport can be included by using the ion heat diffusivity $\chi_i$ to independently normalize its contribution 
relative to the turbulent contribution \cite{angioni_gyrokinetic_2011}:
\begin{equation}
\frac{R}{L_{nB}} = - \frac{RV_B^{\rm GK}/\chi_i^{\rm GK} + R V_B^{\rm Neo}/\chi_i^{\rm an}}{D_B^{\rm GK} /\chi_i^{\rm GK} + D_B^{\rm Neo}/\chi_i^{\rm an}}
\label{eq.nc_combine}
\end{equation}
where $\chi_i^{\rm GK}$ is the ion heat diffusivity calculated in the quasi-linear gyro kinetic simulation and
\begin{equation}
V_B^{\rm GK} = D_B^{\rm GK} \left (C_T {R \over L_{T_B}} + C_u u^\prime_B + C_p \right).
\end{equation}
The anomalous ion heat diffusivity $\chi_i^{\rm an} = \chi_i^{\rm PB} - \chi_i^{\rm NC}$ is the difference between the power balance and neoclassical ion heat diffusivities computed using the transport code {\sc transp} \cite{pankin_tokamak_2004}, in which the neoclassical ion heat diffusivity $\chi_i^{\rm NC}$ is calculated internally by {\sc nclass} \cite{houlberg_bootstrap_1997}. 
The ion heat diffusivity is chosen as the normalizing factor in preference to the total heat diffusivity because it avoids the need to calculate the radiation losses in the power balance analysis. 

\begin{figure}
\centering
\includegraphics[width=95mm]{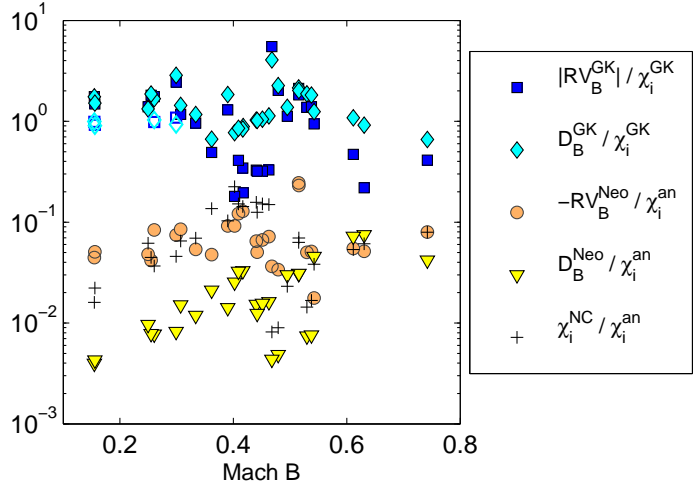}
\caption{Relative importance of neoclassical and turbulent contributions to boron transport at $r/a = 0.5$, comparing each of the terms in Eq. \ref{eq.nc_combine}, and also compared to the ratio of turbulent to neoclassical ion diffusivity.
The quasilinear turbulent contributions are calculated for the $k_\theta \rho_i = 0.3$ mode using $R V_B^{\rm GK} = D_B^{\rm GK}(u^\prime_B C_u + (R/L_T) C_T +C_p)$.  Superscript abbreviations indicate the code used to calculate each quantity [{\sc {\sc gkw}}, {\sc transp} (an, inc. {\sc nclass}), {\sc neo}]. The four TEM points are also shown with an alternative $\chi^{\rm GK}_{i+e}$ normalizing factor (open quadrilaterals). }
\label{fig.neoclassics}
\end{figure}
The neoclassical $RV^{\rm Neo}_B$ is computed using the drift kinetic neoclassical code {\sc neo} \cite{belli_kinetic_2008,belli_eulerian_2009,belli_full_2012}.  Dimensional units are used to build the ratio $RV^{\rm Neo}_B / \chi_i^{\rm an}$, and as an independent cross-check, {\sc neo} finds $RV^{\rm Neo}_B / \chi_i^{\rm Neo}$ between 0 and 4, with values consistent with the dimensionless comparison using the relation $RV^{\rm Neo}_B / \chi_i^{\rm an} = (RV^{\rm Neo}_B / \chi_i^{\rm Neo}) / (\chi_i^{\rm NC} / \chi_i^{\rm an})$.  The comparison of the neoclassical and turbulent contributions to the boron transport is shown in Fig. \ref{fig.neoclassics}. Even though the neoclassical V/D may be quite large, the neoclassical contributions are at least five times smaller than the turbulent ones; when included in the predicted $R/L_{nB}$ using Eq. \ref{eq.nc_combine} the differences are negligible, so for simplicity the neoclassical contribution is omitted in all other figures.  Since the quasilinear gyrokinetic ion heat diffusivity may be under-predicted for TEM simulations (only four of the points), these points are also plotted with $\chi_{e+i}^{\rm GK}$ as the normalizing factor, but this does not alter the result that the turbulent contribution dominates. 

\subsection{Input sensitivities of modelling \label{sec.modelling}}

Even when comparing dimensionless transport coefficients, 
the modelling of such a database requires very well diagnosed plasmas to generate inputs for the simulation.
Since the mode properties largely depend on gradient quantities, very accurate profiles are required
to determine the transport closely enough for a quantitative comparison.
For each point in the database, the inputs to the modelling consist of plasma profiles in
 $T_e$, $T_i$, $n_e$, $v_{\rm tor}$ and the magnetic equilibrium including the $q$ profile.  For each profile, both the local value and the local gradient are an input to the simulation.  The boron rotation is
 used for the main ion toroidal rotation (for all cases, the difference in deuterium Mach number is less than 0.03 using Ref. \cite{kim_neoclassical_1991}).
Each profile fit must be carefully checked, so that bad calibrations and systematic features in the profile shape
can be eliminated.  Cross checking between independent diagnostics can also be useful;
in this database Thomson scattering data was used where available to cross check the $T_e$ profiles 
from electron cyclotron emission (ECE) and the $n_e$ profiles from the IDA inversion.

With this process completed, the most uncertain of the measured inputs to the simulations are the $q$ profile
(which is not well constrained by the magnetics measurements used to generate the equilibria)
and the logarithmic density gradients (as discussed above).  
The temperature profiles are obtained from high resolution localized measurements (ECE and CX) of good quality, 
and do not suffer from the same level of uncertainty.

\begin{figure}
\centering
\includegraphics[width=75mm]{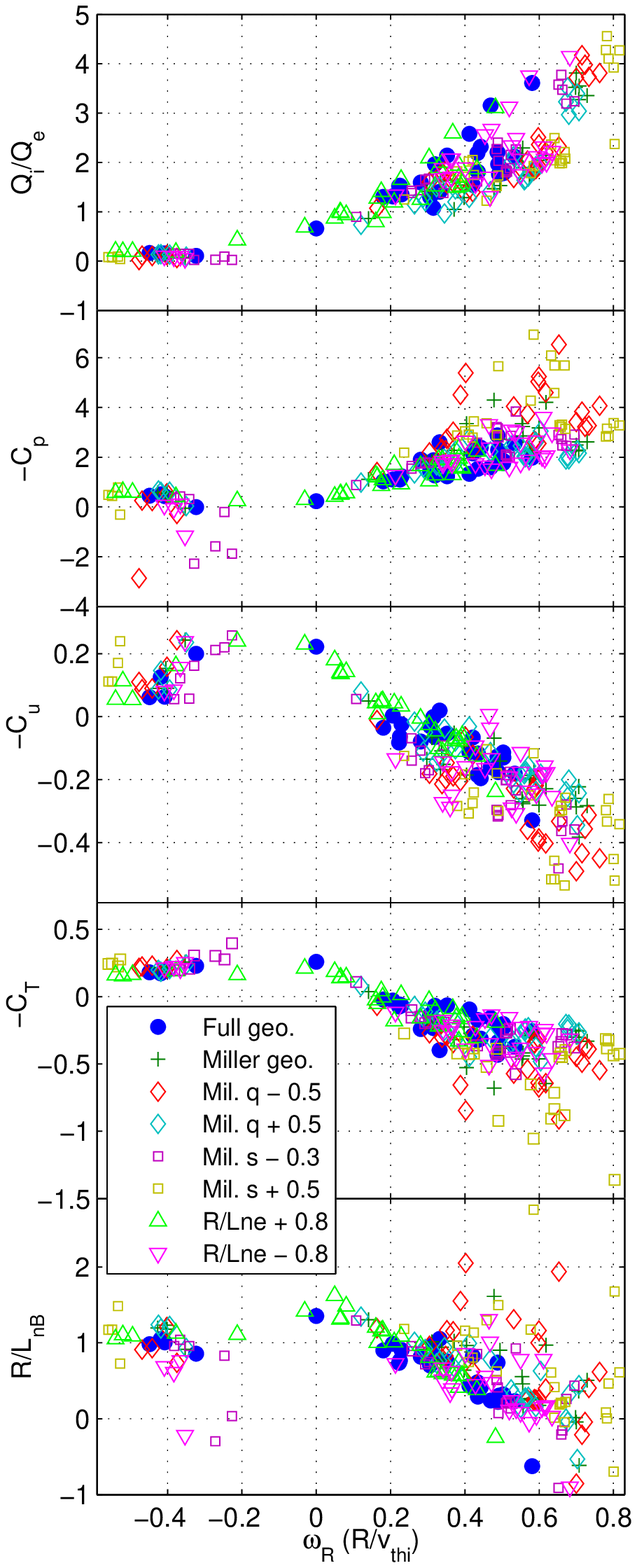}
\includegraphics[width=75mm]{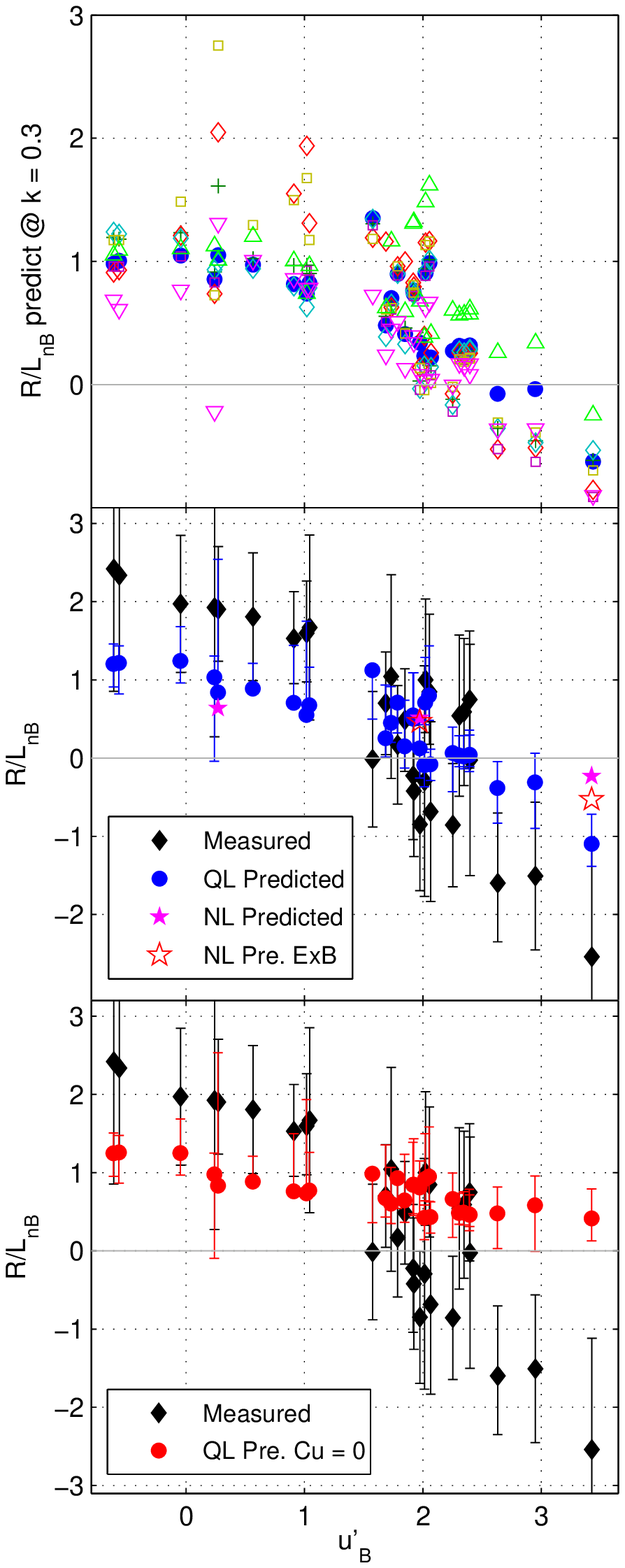}
\caption{\label{fig.sims_jitter} Sensitivity of heat transport and boron transport coefficients to input uncertainties in the safety factor and density profiles for the mode at $k_\theta \rho_i = 0.3$ and $r / a = 0.5$. From top: Ion/Electron heat flux ratio, Convective pinch, rotodiffusion, thermodiffusion, and predicted steady state $R/L_{nB}$ combining the coefficients using measured rotation and temperature gradients. Positive (negative) frequency $\omega_R$ indicates propagation in the ion (electron) diamagnetic direction, signifying an ITG (TEM) mode.}
\caption{Comparison between predicted and measured boron gradients, using the nominal $q$ profiles.  (Top) Sensitivity of predicted $R/L_{nB}$ at $r/a = 0.5$ to the input uncertainties (for the dominant mode at $k_\theta \rho_i = 0.3$).  The points and the legend correspond to those in Fig. \ref{fig.sims_jitter}. (Mid) Comparison of the quasilinear (QL) predicted and measured values of $R/L_{nB}$.  The experimental errors bars are generated as described in the Appendix and are the same as those in Fig. \ref{fig.meas_err}, and the modeling error bars are generated from the sensitivity study of the top figure.  The quasilinear predictions are calculated from a spectrum with its peak at $k_\theta \rho_i = 0.3$, which is compared with nonlinear (NL) simulations for selected points (discussed in Sec. \ref{sec.nonlinear}). 
(Lower) As mid, but excluding the contribution from rotodiffusion. 
\label{fig.compare_sims}}
\end{figure}

To investigate the sensitivity of the modelling results to these uncertainties, a set of simulations
was generated for the entire database, in which the most uncertain input parameters were varied within realistic estimates of their uncertainty (Fig. \ref{fig.sims_jitter}).
The ``nominal result'' set is obtained using the full magnetic equilibria from the AUG function parameterization, while in order to conveniently vary the $q$ profile inputs as
independent parameters, the local geometry Miller parametrization \cite{waltz_ion_1999} is also used (recently implemented in {\sc {\sc gkw}} following the treatment in Ref. \cite{candy_unified_2009}; a benchmark with GS2 \cite{kotschenreuther_comparison_1995} is available \cite{peeters_a.g._gkw_2012}).
Using this geometry, the $q$ inputs were varied by $\pm 0.5$ (to a minimum of 1.1 for stability), and the magnetic shear $s$ was varied by $-0.3,+0.5$ (which precludes any reversed shear configurations).  In addition, the bulk plasma density gradients $R/L_{ne}$ were varied by $\pm 0.8$.  

The results of Fig. \ref{fig.sims_jitter} confirm that the mode frequency at $k_\theta \rho_i = 0.3$ provides a good parameterisation of the particle transport over a large range of plasma conditions.  We note that the main effect of varying the density gradient is to shift all the mode frequencies, with the transport coefficients lying along a single line.  Varying the $q$ profile, by contrast, alters both the frequency and the relationship between the mode frequency and the transport coefficients, increasing the predicted $R/L_{nB}$ most notably for ITG modes with a weak frequency, which were previously closer to the TEM transition.
The largest enhancements to the convective $C_p$ pinch coefficient occur when the alterations to the $q$ profile enhance the parallel compression part of the pinch (the term proportional to $\propto k_\parallel^2 \omega_R$ in Eq. 10 of Ref. \cite{angioni_direction_2006}), which for the ITG reinforces the $E \times B$ curvature pinch ($\propto |\omega|^2$, first in Refs. \cite{garbet_turbulent_2003,baker_perturbative_2004}).  The parallel compression pinch increases as magnetic shear increases and $q$ decreases ($k_\parallel^2 \propto (s / q R)^2$), and following Ref. \cite{angioni_off-diagonal_2012} we refer to it as the ``slab resonance''.   This provides
the counterpart to the previously reported result of increased electron peaking with magnetic shear in the TEM regime \cite{fable_role_2010,angioni_intrinsic_2011}, but we note that that effect is also partially due to
the increase in the curvature drift frequency enhancing the inward thermodiffusion.  These observations are consistent with the impurity transport mechanisms discussed in Ref. \cite{angioni_direction_2006,angioni_non-adiabatic_2007, angioni_off-diagonal_2012}.

The range of possible predicted $R/L_{nB}$ generated by varying the input $q$ profiles and bulk species density gradients is used to generate
an error bar on the `nominal result' for each point (the result obtained using the default experimental profile fits (IDA for density and a function parametrization for the equilibrium).  The sensitivities of the $k_\theta \rho_i = 0.3$ mode are used to assume the error for the quasilinear result including all 5 modes (Fig. \ref{fig.compare_sims}).

\begin{figure}
\centering
\includegraphics[width=50mm]{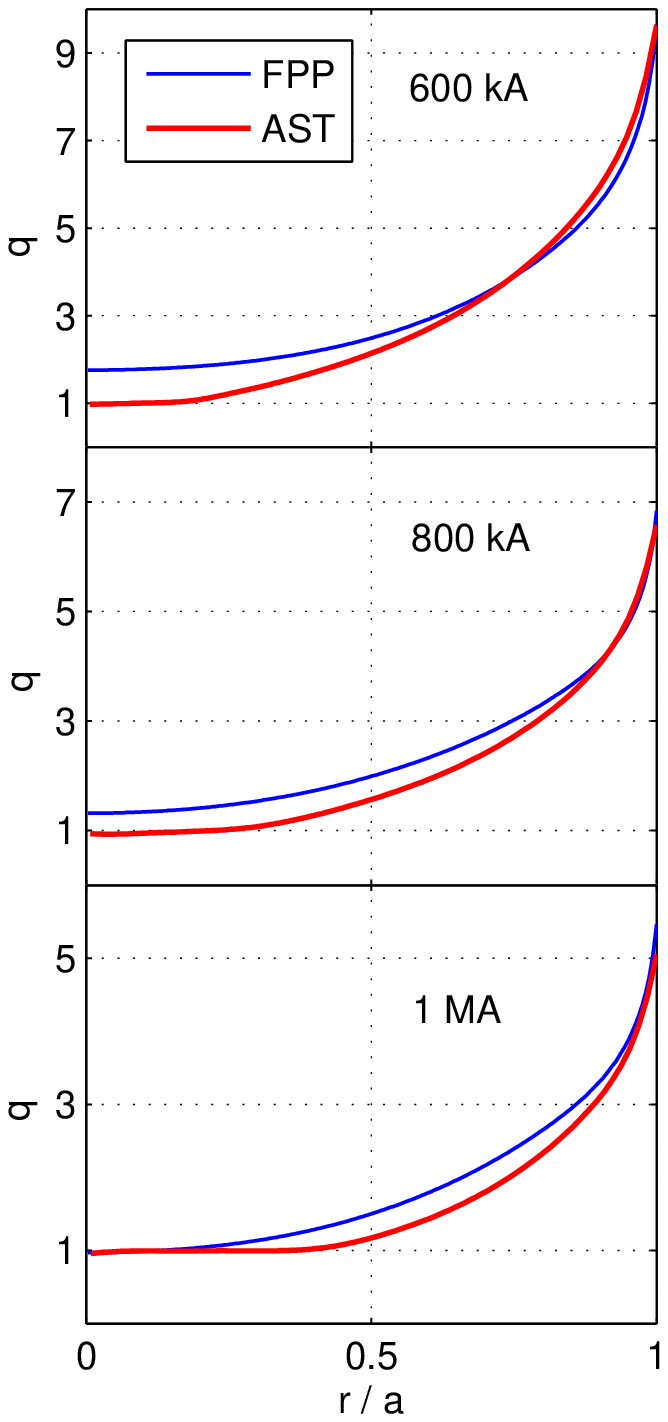}
\includegraphics[width=70mm]{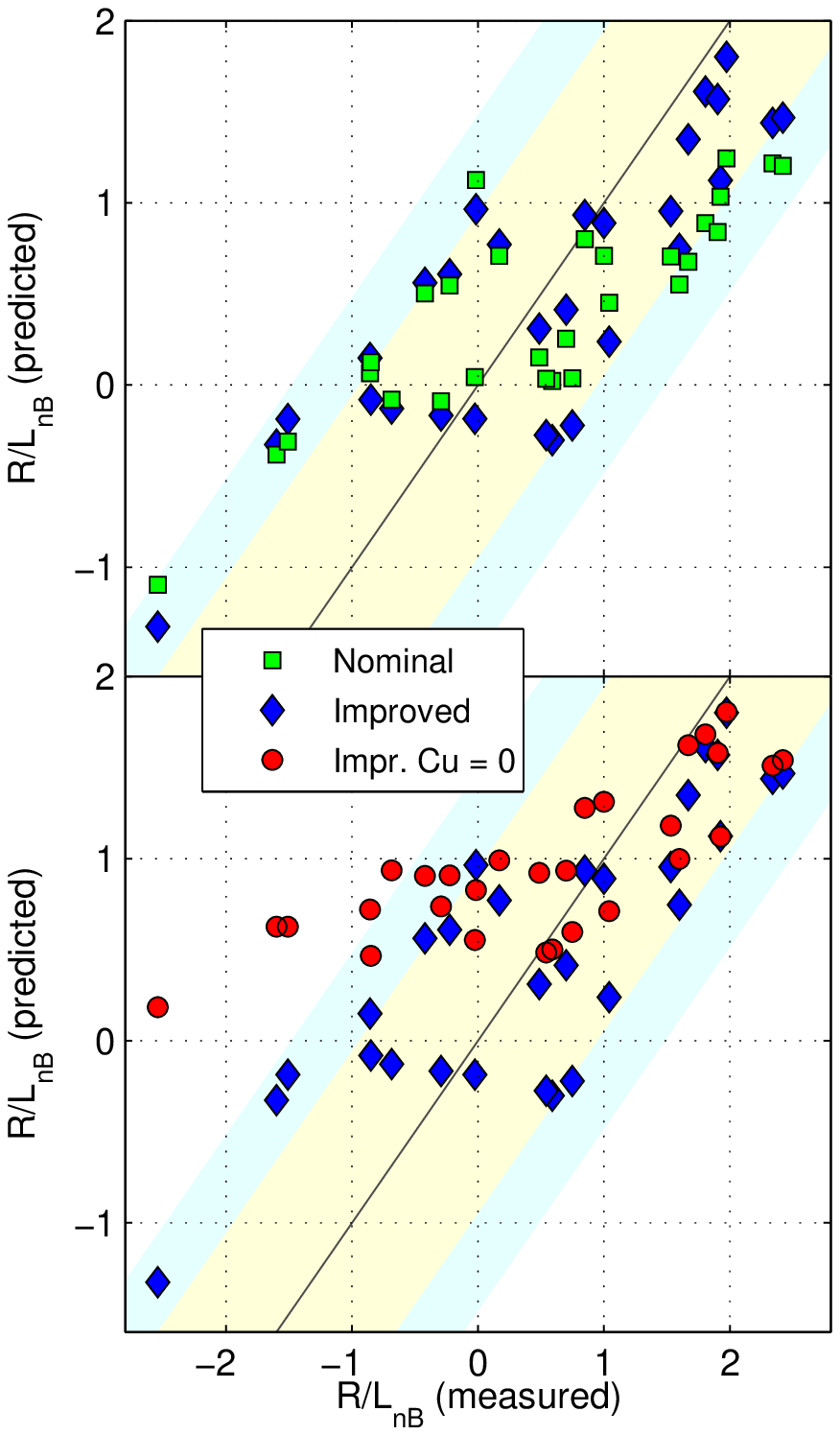}
\caption{Comparison of example $q$ profiles from AUG function parametrization and {\sc astra}. \label{fig.q}}
\caption{\label{fig.compare_sims2}(Upper) Comparison between predicted and measured boron density gradients, for the quasilinear simulations with nominal (squares) and improved (diamonds) $q$ profiles with slight ITG mode biasing.
(Lower) Comparison between the improved quasilinear predictions including (diamonds) and excluding (circles) the rotodiffusion effect.
In both plots, the inner shaded band indicates the mean of the experimental uncertainty.  The outer shaded band additionally includes the mean of the variation
found in the simulation sensitivity study (for the nominal profiles).  These bands therefore represent the means of the individual error bar sets shown in Fig. \ref{fig.compare_sims}.}
\end{figure}

Given the sensitivity of the results to the slab resonance for the ITG mode, a further iteration of the sensitivity study is presented, using more realistic $q$ profiles modelled with the {\sc astra} transport code coupled with the equilibrium solver {\sc spider} \cite{fable_progress_2012}.  Each shot is simulated in interpretive mode over a time window covering all the phases in the database.  The current profile is allowed to evolve from an initial condition of $(\partial / \partial r)(R E_\phi) = 0$, including {\sc torbeam} \cite{poli_torbeam_2001} for EC current drive, and a simple sawtooth model to fix the profile inside the inversion radius.  The resulting $q$ profiles show good agreement with the measured inversion radius, and are the most accurate possible in the absence of a direct measurement.  The comparison of these $q$ profiles with those of the function parametrization in Fig. \ref{fig.q} indicates reduced $q$ at mid radius, and reduced shear for the lower current cases (both conducive for the slab resonance).  In all equilibria, the positions of the flux surfaces are much better constrained than the $q$ profile, such that switching between the nominal and improved equilibria for profile mappings has a negligible impact on the inputs for the simulations.

The simulation set with the improved $q$ profiles has also been altered slightly to favour the ITG mode, relative to the nominal simulation parameters.    For the 4 points which previously had TEM at $k_\theta \rho_i = 0.3$, $Z_{\rm eff}$ is increased by 0.1, two points have $R/L_n$ decreased by 0.4, and one has $R/L_{Ti}$ increased by 0.4.  In addition, the more physical friction and energy scattering terms in the collision operator have been added for the entire set.
The resultant switch from TEM dominance at the largest scales is necessary for the 4 previously TEM dominated cases to find sufficient levels of ion heat transport in comparison to the power balance calculation described in Sec. \ref{sec.nonlinear}.  These slight changes to the inputs are enough to push the dominant modes at $k_\theta \rho_i = 0.3$ to ITG, (some TEM modes persist at higher $k_\theta$), to better match the heat flux ratio, and to allow the slab resonance in the impurity convection to give increased boron peaking for these (lower rotation gradient) cases.  The comparison between the simulations with both sets of $q$ profiles and the measurement is shown in Fig. \ref{fig.compare_sims2}.

\subsection{Comparison of modelling and measurement, interpretation}
  
In this section, the results of the boron measurement and the quasilinear gyrokinetic modelling are compared and interpreted.
In Fig. \ref{fig.compare_sims} the results with the nominal profiles are plotted against the dominant experimental correlation of the plasma rotation gradient, including the error bars of the simulation sensitivity studies (described in the previous sections) and measurement (described in the Appendix). In Fig. \ref{fig.compare_sims2}, the more direct comparison between predicted and measured $R/L_{nB}$ is shown, between the nominal and improved profiles, and with and without rotodiffusion.  The shaded uncertainty bands represent the 
mean uncertainty of the experimental density gradients, and the mean of the simulation sensitivities discussed in the previous section. 

Three main conclusions can be immediately
drawn from these comparisons:  First, both sets of modelling reproduce the experimental correlation, with the strength of the correlation sensitive to the $q$ profile and the dominant mode.
Second, nearly every point can be matched quantitatively within the estimated uncertainties, with the best agreement
found for the improved $q$ profiles and dominant ITG modes, which allow the most peaked cases to find the ITG slab resonance.   Third, the modelling is only able to obtain the most hollow boron profiles in agreement with
the measurements (these are the points at higher rotation gradients) if the rotodiffusive contribution to the boron transport is included.  The quantitative agreement for the most hollow profiles is not perfect, an exact match would require either the rotodiffusive or thermodiffusive component of the transport to be roughly a factor of two larger (see Sec. \ref{sec.sym}).

Given the demonstrated sensitivity of the modelling to the inputs, the combination of fluxes of opposite signs, and the sensitivity of the measurement to the deuteron profile, 
we find this agreement to be quite remarkable, building confidence that the local gyrokinetic model is correctly describing the dominant influences on the impurity transport.

The success of the modelling allows us to understand the dominant experimental anti-correlation of boron peaking with the rotation gradient as a
combination of correlations which connect the turbulence and the heating systems.  The connections between the transport channels are framed
in terms of the mode frequency, which we have demonstrated is the ubiquitous parameter for turbulent particle transport. 

\begin{figure}
\centering
\includegraphics[width=70mm]{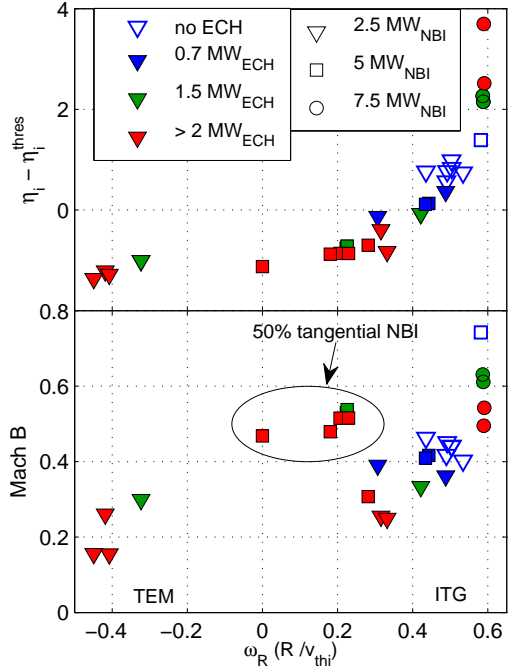}
\caption{(Upper) The mode frequency at $k_\theta \rho_i = 0.3$ is strongly related to the distance from the analytic $\eta_i = L_n / L_T $ threshold for the ITG mode \cite{guo_linear_1993}, where $\eta_{\rm thres} = 1.0 + \sqrt{1.0 + (1 + T_i/T_e)(L_n/R q k_\theta \rho_i)^2}$ is derived in circular geometry. (Lower) Correlation between the mode frequency and boron Mach number is related to the AUG auxiliary heating systems (represented by the same symbol colors and shapes as in Fig. \ref{fig.scatter}).}
\label{fig.thres}
\end{figure}

The most fundamental relationship is that the auxiliary heating systems directly determine the temperature profiles, which set the characteristic frequency of the turbulence. For ITG modes this can be seen in the direct relationship between the mode frequency and $\eta_i = L_n / L_T$ shown in Fig. \ref{fig.thres}.  However, 
it can be more helpful to view this as a flux-driven relationship:  The ratio of ion and electron turbulent heat fluxes $Q_i/Q_e$ must match the power balance.  This ratio has a direct relationship with the mode frequency, as demonstrated in Fig. \ref{fig.sims_jitter}.

Since the turbulence dominates all transport channels, the selection of the mode frequency by the power balance has an immediate
consequence for the electron density profile, in a relationship that is now well understood \cite{angioni_density_2003,fable_role_2010,angioni_off-diagonal_2012}.
In the set of NBI heated, ITG dominated H-modes in the present dataset, this relationship manifests itself as an increased density peaking when electron heating is applied (as shown in Fig \ref{fig.example}, and Ref. \cite{mcdermott_effect_2011}). The density peaking increases as the electron heating
shifts the turbulence from strong ITG towards the TEM transition (in the frequency domain), which is the frequency region with the strongest convective particle pinch \cite{fable_role_2010,angioni_off-diagonal_2012}.  Of course, the turbulence is not determined by a single linear mode, so the transition from ITG to TEM is a gradual one, and feeds back on the heat flux relationship:  Increased electron heating increases the prevalence of TEM even when ITG modes dominate; this increases the density peaking, which reduces $\eta_i$ and, therefore, the mode frequency further, a process that reinforces itself until the balance between ITG and TEM that matches the energy fluxes is reached. 

The rotation and its gradient also have a more circumstantial correlation with the mode frequency (Fig. \ref{fig.thres}), 
The NBI heating directly drives both rotation and ion temperature gradients (which affect the mode frequency through an increase of $\eta_i$).
It can also be seen in Fig. \ref{fig.thres} that this relationship depends on the beam geometry, 
since the most tangential NBI beam drives relatively more rotation for less increase in $R/L_{Ti}$ at mid radius.
In more flexible geometry configurations, such as two oppositely directed NBI beams (as in DIII-D \cite{politzer_influence_2008}), this relationship could be broken altogether.
Electron heating drives TEM, with the influence on the frequency and the electron peaking as discussed above (which also reduces $\eta_i$).  The increased
density peaking has a consequence for the residual stress \cite{angioni_intrinsic_2011}, which 
acts to reduce the rotation (Fig \ref{fig.example} and Ref. \cite{mcdermott_effect_2011}).

Invoking the above connections between the rotation, ion and electron temperature profiles of the bulk plasma, we 
see that the anti-correlation between boron peaking and rotation gradient, is a combination of two effects:  First,
the (partly circumstantial) correlation between rotation and mode frequency (Fig. \ref{fig.thres}) combines with the
direct connection of the mode frequency to the impurity peaking (Fig. \ref{fig.sims_jitter}) 
to produce a weak correlation that can be seen in the lower plot of Fig. \ref{fig.compare_sims} in which the rotodiffusive contribution $C_u u^\prime_B$ is excluded. 
Second, the rotodiffusive contribution to the boron flux becomes important at higher rotation gradients, as is clear
from the comparison with the middle plot in Fig. \ref{fig.compare_sims} which keeps the rotodiffusion.  Therefore 
both the correct turbulence frequency and the rotodiffusion are required together to explain the observed anti-correlation.   
We also note that in the cases
in which the relationship between frequency and rotation is weakest (due to the use of the more tangential beams), the
rotodiffusive contribution is stronger, compensating the other terms to explain the near linear observed correlation.
Since, as a trace species, boron has no back reaction on the turbulence, its transport can be
considered an independent diagnostic of the turbulence theory.  This is in contrast to the bulk transport channels
in which the driving gradients must be necessarily consistent with the fluxes.

In fact, the anti-correlation between boron peaking and rotation gradient is the most complex relationship in the database to describe, and the framework above 
also describes the other correlations visible in Fig. \ref{fig.scatter}.  The temperature gradients (and heating powers) have direct relationships with
the mode frequency, which also connect to the density gradient.  In addition, the boron density gradients are correlated with the electron density gradients because the same curvature pinch mechanism applies to all species.   For impurities, the roto- and thermodiffusive terms reduce the boron density peaking relative to the electron density peaking (for electrons, rotodiffusion is negligible, and thermodiffusion has opposite sign) \cite{angioni_off-diagonal_2012}.   

\section{Nonlinear modelling \label{sec.nonlinear}}

In this section we verify the quasilinear (QL) modelling against nonlinear (NL) simulations for three representative
points from the database, at $u^\prime_B= 0.27$, $u^\prime_B=1.98$ and $u^\prime_B=3.43$ (with the nominal q profiles) as shown in Fig. \ref{fig.compare_sims}.  As in the
QL cases, each simulation contains four trace species to enable independent determination of each
of the dimensionless boron transport coefficients.  The nonlinear simulations
are well resolved for energy fluxes (21 toroidal modes, 167 radial modes, 
 25 parallel grid points and 32 x 8 velocity grid points), but we have not excluded that the complete convergence in
 the impurity channels might require greater resolution. 

\begin{figure}
\includegraphics[width=150mm]{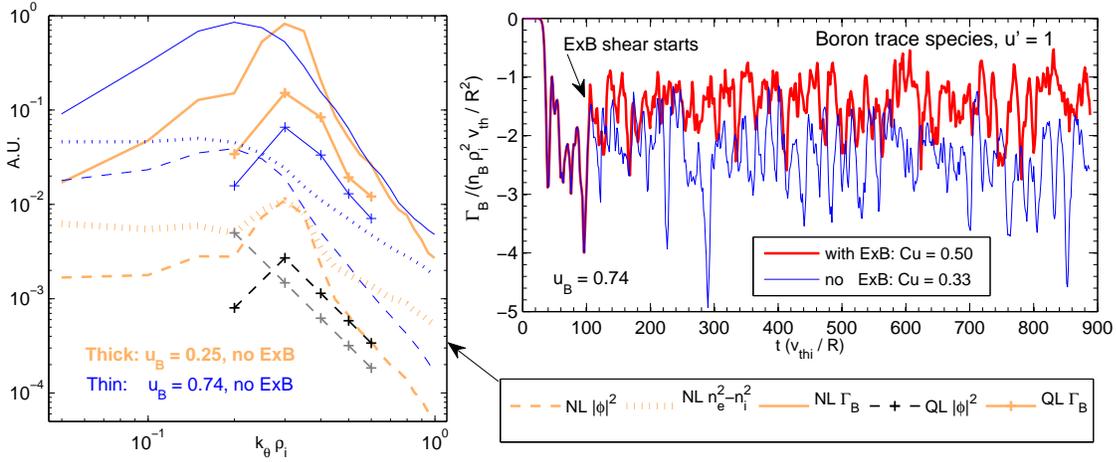}
\caption{(Left) Comparison of nonlinear and quasilinear spectra of electrostatic potential, vorticity, and boron flux for species with no gradients for the $u=0.25$ and $u=0.74$ cases.  The dashed lines with crosses represent the two spectra used for the quasilinear weightings. (Right) Nonlinear particle flux for the trace boron species with $\{R/L_T,R/L_n,u^\prime_B\}=\{0,0,2.31\}$ for the $u=0.74$ case, with and without $E \times B$ shear coupled for pure toroidal rotation \cite{casson_anomalous_2009}.  For this trace, $\Gamma_B = n_B D_B(2.31 \cdot C_u + C_p)$.  The two curves are coincident until the $E \times B$ flow begins at $t=100(R/v_{thi}$).}
\label{fig.nonlinear}
\end{figure}

Each of the selected points have dominant ITG turbulence at larger scales, but the lower rotation cases 
have an increasing TEM character at smaller scales.  It can be seen from the spectra that increased TEM
activity shifts the spectral peak to smaller scales, and changes the shape of the spectra (Fig. \ref{fig.nonlinear}). 
The nonlinear simulations produce self-consistent spectra through a turbulent cascade, whereas 
the spectra used to weight the wavenumbers in the QL simulations are prescribed.  
For the QL weighting, the spectral shape of Ref. \cite{casati_validating_2009} is used,
with the location of the spectral peak fixed close to that found in the NL simulation (as shown in Fig. \ref{fig.nonlinear})
to enable efficient simulation over a large database (in Ref. \cite{casati_validating_2009} the peak location is adapted 
dynamically depending on the linear growth rates).

For the two lower rotation cases, 
it can be seen in Fig. \ref{fig.compare_sims}  that the differences between the NL and the QL results are no more significant
than the input sensitivities of the simulation. The QL spectral peak is in good agreement with the NL
one (Fig. \ref{fig.nonlinear}), and each of the impurity transport coefficients (Table \ref{tab.nonlinear}) and predicted $R/L_{nB}$ match very closely 
(within $\pm 0.03$ for $C_u$, $\pm 0.12$ for $C_T$, and $\pm 0.3$ for $C_p$).
The heat flux ratios match less closely, because the QL spectrum used 
does not contain enough long wavelength modes which dominate the heat fluxes more.

For the highest rotation gradient $u^\prime_B=3.43$, strongest ITG case, the spectral mismatch only partly explains the small discrepancy
between the QL and NL results.  In Table \ref{tab.nonlinear}, the QL result is re-weighted with the spectral peak 
at $k_\theta \rho_i = 0.2$ (dashed grey in Fig. \ref{fig.nonlinear}), which gives excellent agreement between $C_u$ and $C_p$, but a remaining discrepancy
in $C_T$ of 0.07.  Since the $R/L_{nB}$ result depends on this value multiplied by $R/L_{Ti}$, (= 5.8 for this case),
accurate determination of the hollowest profiles requires an extremely precise determination of $C_T$.
It can be seen that from the single mode values that $C_T$ (and $C_u$) increase strongly at high $k_\theta$ for this case, 
so the exact spectral slope as well as the peak becomes important.

As a cross-check on the simulations, the ion-electron heat flux ratios are also compared with the value from 
a power balance (PB) calculation in which the NBI ion/electron heating and equipartition power within $ r/a = 0.5$
is computed by {\sc transp}, and the radiation losses are assumed to be a fixed 50\pc of the auxiliary heating power on the electrons within $r/a = 0.5$, and where all the ECRH power is centrally deposited.
Since this ratio is quite sensitive to the approximate radiation model, this quantity should only be treated as a qualitative check 
that the simulations are obtaining the relevant type of turbulence; indeed the values shown in Table \ref{tab.nonlinear} confirm that this is the case.

For the four cases (for the nominal parameters) with dominant TEM even at the largest scales, nonlinear simulations find $Q_i/Q_e$ much less than the power balance value (but similar to the QL result), which provides the motivation for the ITG biased set in Fig. \ref{fig.compare_sims2}.  The nonlinear TEM runs are also much more sensitive to high-$k_\theta$ energy pileup most evident in the vorticity spectrum \cite{krommes_role_1994,scott_computation_2006}, and we have not yet found the correct dissipation to avoid this problem.  In the attempted TEM simulations a saturated state is achieved with heat fluxes, but the particle fluxes are more sensitive to these numerical difficulties, and give unphysical results far from the QL results.  The ITG cases shown here are more robust, and need only a small amount of $k^4$ hyper-dissipation to generate a clean vorticity spectrum (Fig. \ref{fig.nonlinear}, defined here as ${\tilde n}_e^2-{\tilde n}_i^2$) with robust particle fluxes.

In summary, the quasilinear results using only the dominant mode at each scale, reproduce the nonlinear results for particle transport 
in ITG dominated turbulence down to a very precise level for every coefficient (when the correct spectral shape is used in the quasilinear weighting).  
In regions of transitions between turbulence regimes both the quasilinear and nonlinear simulations are less robust, making comparisons much more sensitive.

\begin{table}
\caption{Comparison of dimensionless transport coefficients between nonlinear simulations, the quasilinear spectrum, and the quasilinear value for each individual mode. The three cases correspond to the three nonlinear points in Fig. \ref{fig.compare_sims}, which use the nominal $q$ profiles.  Nonlinear values are time averaged for at least 300 $(R / v_{thi})$ after saturation, and different averaging windows do not change any of the coefficients by more than 5\pc.  All modes are ITG except where indicated TEM.\label{tab.nonlinear}}  
\begin{indented}
\begin{scriptsize}
\lineup
\item[]\begin{tabular}{@{}lrrrrrrr}
\br
Case &   $C_u$   &   $C_p$   &  $C_T$   & $\frac{R}{L_{nB}} $  & $\frac{\chi_i}{\chi_e}$ &  $ \frac{Q_i}{Q_e} $ &  $\frac{{Q_i}^{\rm pb}}{{Q_e}^{\rm pb}}$ \\
\br
\qquad $u^\prime_B = 0.27$, $u_B = 0.25 $,  $R/L_{Ti} = 3.92 $, $R/L_{Te} = 8.23 $  \\
NL                            &    0.02 &   -2.20 &      0.40 &      0.64 &      2.64 &      1.26 &  1.02 \\
QL (spectrum 0.3)             &   -0.02 &   -2.39 &      0.40 &      0.84 &      1.29 &      0.61 &  \\
QL (spectrum 0.2)             &   -0.03 &   -2.34 &      0.33 &      1.07 &      2.13 &      1.01 &  \\
$k_\theta \rho_i = 0.2$       &   -0.04 &   -2.31 &      0.28 &      1.23 &      3.73 &      1.78 &  \\
$k_\theta \rho_i = 0.3$       &   -0.02 &   -2.60 &      0.40 &      1.05 &      2.95 &      1.41 &  \\
$k_\theta \rho_i = 0.4$       &    0.00 &   -3.38 &      0.68 &      0.72 &      2.41 &      1.15 &  \\
$k_\theta \rho_i = 0.5$ (TEM) &   -0.13 &    3.29 &     -0.39 &     -1.74 &      0.03 &      0.01 &  \\
$k_\theta \rho_i = 0.6$ (TEM) &   -0.16 &    2.79 &     -0.42 &     -1.12 &      0.03 &      0.01 &  \\
\mr 
\qquad $u^\prime_B = 1.98$, $u_B = 0.50 $, $R/L_{Ti} = 4.70 $, $R/L_{Te} = 6.61 $ \\
NL                       &      0.11 &   -1.73 &      0.21 &      0.51 &      3.39 &      2.41 &   2.40 \\
NL with ExB              &      0.25 &   -1.88 &      0.20 &      0.47 &      3.36 &      2.39 &   \\
QL (spectrum 0.3)        &      0.14 &   -1.43 &      0.11 &      0.63 &      2.72 &      1.93 &    \\
QL (spectrum 0.2)        &      0.06 &   -1.23 &      0.06 &      0.81 &      2.88 &      2.05 &    \\
$k_\theta \rho_i = 0.2$  &      0.00 &   -1.05 &      0.02 &      0.96 &      3.02 &      2.14 &    \\
$k_\theta \rho_i = 0.3$  &      0.06 &   -1.26 &      0.07 &      0.82 &      2.76 &      1.96 &    \\
$k_\theta \rho_i = 0.4$  &      0.16 &   -1.52 &      0.13 &      0.60 &      2.67 &      1.90 &    \\
$k_\theta \rho_i = 0.5$  &      0.29 &   -1.82 &      0.21 &      0.23 &      2.54 &      1.81 &    \\
$k_\theta \rho_i = 0.6$  &      0.50 &   -2.26 &      0.35 &     -0.38 &      2.31 &      1.64 &    \\
\mr
\qquad $u^\prime_B = 3.42$,  $u_B = 0.74 $, $R/L_{Ti} = 5.80 $, $R/L_{Te} = 4.70 $ \\
NL                       &      0.33 &   -1.98 &      0.19 &   -0.23 &      2.65 &      2.98 &   3.78 \\
NL with ExB              &      0.50 &   -2.08 &      0.16 &   -0.53 &      2.50 &      2.82 &   \\
QL (spectrum 0.3)        &      0.44 &   -2.27 &      0.32 &   -1.10 &      2.72 &      3.34 &   \\
QL (spectrum 0.2)        &      0.33 &   -2.01 &      0.26 &   -0.62 &      3.06 &      3.76 &   \\
$k_\theta \rho_i = 0.2$  &      0.25 &   -1.83 &      0.22 &   -0.28 &      3.39 &      4.17 &   \\
$k_\theta \rho_i = 0.3$  &      0.33 &   -1.98 &      0.26 &   -0.62 &      2.93 &      3.61 &   \\
$k_\theta \rho_i = 0.4$  &      0.48 &   -2.36 &      0.33 &   -1.22 &      2.60 &      3.20 &   \\
$k_\theta \rho_i = 0.5$  &      0.77 &   -3.09 &      0.51 &   -2.50 &      2.21 &      2.71 &   \\
$k_\theta \rho_i = 0.6$  &      1.58 &   -5.17 &      1.06 &   -6.35 &      1.72 &      2.12 &   \\
\end{tabular}
\end{scriptsize}
\end{indented}
\end{table}

\subsection{Symmetry breaking and rotodiffusion \label{sec.sym}}

The rotodiffusive contribution to the particle transport, $C_u u^\prime_B$,  is driven by the gradient $u^\prime_B$ of the impurity rotation.   
The mechanism is a consequence of symmetry breaking mechanisms in the local gyrokinetic equation \cite{peeters_overview_2011}, such that the value of the coefficient $C_u$ depends on both the mode frequency (Fig. \ref{fig.sims_jitter}), the rotation, and the nonzero $k_\parallel$ developed by a mode due to the symmetry breaking ($ C_u \propto 2 u - {(Z A_B/2A_D)(k_\parallel v_{thi}/ \omega_D)}$)  \cite{camenen_impact_2009}.  The combination of these two terms means that there is no special significance to $C_u=0$ seen in Fig. \ref{fig.sims_jitter}.  By performing numerical experiments with either the rotation or the rotation gradient of the bulk plasma removed, we have verified that the $k_\parallel$ generated by the $u^\prime$ symmetry breaking (of the bulk species) is dominant over the $u$ contribution to $C_u$ in this dataset (also true in the cases in Ref. \cite{camenen_impact_2009}).
   Each symmetry breaking mechanism that can generate
a momentum flux can also generate rotodiffusion.  In our quasilinear simulations, three possible symmetry breaking mechanisms
contribute to the rotodiffusion:  The gradient in the bulk species rotation (responsible for diagonal momentum transport \cite{peeters_linear_2005}), the Coriolis force (which generates a momentum pinch \cite{peeters_toroidal_2007}), and the up-down asymmetry of the flux surfaces \cite{camenen_transport_2009}.  The last of these is not present when the Miller geometry is used, and only has a very minimal effect at the $r/a=0.5$ location  studied.   Symmetry breaking by $E \times B$ shear can also be considered in the local model \cite{casson_anomalous_2009}, but as a coupling of many modes is only meaningful in nonlinear simulations.  

To assess the impact of this additional symmetry breaking mechanism, the two nonlinear cases at $u_B=0.5$ and $u_B=0.74$ were repeated including the $E \times B$ shearing that is required to produce a purely toroidal rotation gradient (in the cases without $E \times B $ shear, only the parallel projection of the rotation gradient is kept).  The results, shown in Figs. \ref{fig.compare_sims} and \ref{fig.nonlinear} and Table \ref{tab.nonlinear}, demonstrate the $E \times B$ shear can increase the rotodiffusion coefficient by up to 0.4 in the high rotation cases.  This result demonstrates for the first time that each symmetry breaking mechanism that is relevant to momentum transport has its counterpart in the rotodiffusive particle transport, and that we have understood well the connections
between these off-diagonal transport mechanisms \cite{angioni_off-diagonal_2012}.  Tests conducted by switching the sign of the plasma current in the simulation geometry indicate that for purely toroidal rotation, the $E \times B$ contribution to the rotodiffusion always acts to increase the rotodiffusion coefficient (for momentum transport this contribution always acts to reduce the effective momentum diffusivity \cite{casson_anomalous_2009}).
 
In a global model, other symmetry breaking mechanisms can be present, and while formally smaller in terms of a $\rho_* = \rho_i / L_\perp $ small parameter expansion \cite{peeters_overview_2011}, are thought to be important in determining the non-negligible turbulent residual stress \cite{angioni_intrinsic_2011}.  A leading candidate for relevant global symmetry breaking mechanisms is the profile shearing effect \cite{gurcan_residual_2010,camenen_consequences_2011,waltz_gyrokinetic_2011}.  The inclusion of these effects via global simulations might be expected to have a further influence on the rotodiffusive term, and is expected to increase 
the rotodiffusion coefficient in the ITG regime (in parallel to the inward directed
contribution to the residual stress \cite{angioni_intrinsic_2011}), which would improve the agreement for the most hollow $R/L_{nB}$ points at the highest Mach number. 

\section{Conclusions}

In this work, high quality charge exchange measurements have been used to construct a 
database of boron density profiles for a range of H-mode plasmas in the ASDEX upgrade tokamak.  
In addition to better measurements, this database improves on previous work \cite{angioni_gyrokinetic_2011}, 
by covering a larger range of auxiliary heating conditions, $q$ profiles, and plasma rotation.
This improved boron database allows quantitative comparison with modelling of the predicted
logarithmic gradient of the boron density at mid radius, where neoclassical transport of boron is demonstrated
to be negligible.  The ability to perform quantitative comparisons represents a significant advance over the previous
qualitative study, since it enables delicate conclusions to be drawn about the precise off-diagonal transport mechanisms at play.

Quasilinear gyrokinetics is used to independently predict the turbulent contribution, by computing 
the ratio between turbulent diffusion and off-diagonal pinch, thermodiffusive and rotodiffusive components.
For a subset of points, the quasilinear approach is validated against nonlinear simulations 
and is shown to closely reproduce the relative contribution of each off-diagonal component. 
To enable robust quantitative comparison, the effect of uncertainties in both the measurement 
and simulation inputs have been investigated.  For the simulated dataset, the density profile and 
$q$ profile have the largest input uncertainties, both of which influence the predicted boron gradient.
Changes in the input density and temperature gradients change the mode frequency and thus the particle transport, 
while changes in the $q$ profile alter the contributions between parallel (slab) and perpendicular compression (curvature) 
resonances in the convective pinch for the ITG mode.  These sensitivities highlight the 
need for excellent diagnostics and equilibrium reconstructions when making quantitative comparison 
with gyrokinetic modelling.  Accurate $q$ profiles are particularly vital for correct modelling of impurity
 transport channels due to the toroidal and slab resonances in the convective term.

Within the investigated uncertainties, the modelling is able to quantitatively
match the measured boron gradients, and reproduces the correlations
observed in the experimental database.  The primary correlation in the database, between 
boron density peaking and rotation gradient, is explained by the modelling as due to a combination of 
the turbulence regime and an impurity flux driven by rotation gradient (the rotodiffusion effect), both of 
which are indirectly determined (and correlated) by the auxiliary heating power.

The rotodiffusive contribution becomes important 
at the higher rotation gradients in the dataset, and is required to reproduce hollow profiles 
in agreement with the measured values.  This constitutes an 
experimental confirmation of rotodiffusion, which is a consequence of the same symmetry breaking mechanisms which drive
momentum transport.  In nonlinear simulations, the addition of symmetry breaking by background perpendicular flow shear increases
the rotodiffusive mechanism, and we speculate that further mechanisms of residual stress might
improve agreement further in global simulations.  Centrifugal effects
were included in these simulations but do not significantly affect the results for the boron impurity at mid radius.

In the absence of a direct measurement, the most accurate $q$ profiles are 
obtained from an interpretive transport simulation coupled to an equilibrium solver. These 
$q$ profiles allow the simulations to correctly determine the ITG parallel compression (slab) resonance in the impurity pinch,
and give the best agreement between the boron predictions and measurement.  The slab resonance is most
significant for the more peaked boron profiles (the lower rotation cases), which are also nearest to the ITG / TEM transition.

The strong correlation of all the boron transport coefficients with the turbulent mode frequency underlines its convenience
as a ubiquitous parameter for characterizing turbulent particle transport.  
Since the ratio of ion and electron fluxes is also determined by the mode frequency and must match the power balance, 
 simulations that match all transport channels consistently provide the most robust validation of 
the turbulent transport models.  Indeed, in the context of uncertain inputs, iterating in the parameter space 
to match multiple transport channels could even be used to further constrain the measurement uncertainties.
We believe the results presented in this work constitute a subtle and stringent test of the 
local gyrokinetic model, and build confidence in its predictive power with respect to the transport of light impurities.

\vspace{10 mm}

\textbf{Acknowledgements}:  
We are indebted to all those who have worked on the 
{\sc chica, fidasim, {\sc gkw}, gs2, cliste, ges, neo, transp, nubeam, nclass, astra, spider, torbeam, chease, cview, augped, rwshot, sf2ml}, and {\sc ida} codes, 
all of which contributed directly or indirectly to this work.  
In particular F.J.C. would like to thank A.G. Peeters and E. Belli for making the {\sc {\sc gkw}} and {\sc neo} codes freely available.
Simulations were performed on the {\sc hpc-ff} (project {\sc wimp}) and {\sc helios} (project {\sc residual}) machines, 
and using the resources of the Rechnungzentrum at Garching.  {\sc transp} simulations were performed at {\sc pppl}.

\section{Appendix: Estimation of measured gradient uncertainties}

The sample errors bars shown in Fig. \ref{fig.example} are the mean error bars provided by the {\sc chica} code and include the CX fitting error, in addition to systematic uncertainty estimates for the atomic data and intensity calibration. 
One further uncertainty in the calculation is the deuteron density profile
required for the {\sc fidasim} modelling of the neutral contributions, a quantity not routinely measured on AUG.  The comparison
of the {\sc fidasim} modelling with beam emission spectroscopy in Ref. \cite{dux_impurity_2012} demonstrated that the modelling of the 
halo contribution to the CX signal is accurate to within 30\pc{}, part of the discrepancy could be due to uncertainties in the deuteron density.

The deuteron profile must be estimated from the electron density profile combined with an assumption about the overall impurity concentration.  In addition, the profiles of electron density (primarily derived from an inversion of 5 horizontal line integrated interferometry measurements) also contribute to the uncertainty.  The electron profiles used in this database are obtained from an inversion of the interferometry including the edge lithium beam, obtained from the Bayesian integrated data analysis (IDA) tool \cite{fischer_probabilistic_2008,fischer_integrated_2010}. 
The deuteron density influences the beam penetration and attenuation, and the relationship between the deuteron profile uncertainty and the calculated boron density cannot be surmised a priori. 

\begin{figure}
\begin{center}
\includegraphics[width=120mm]{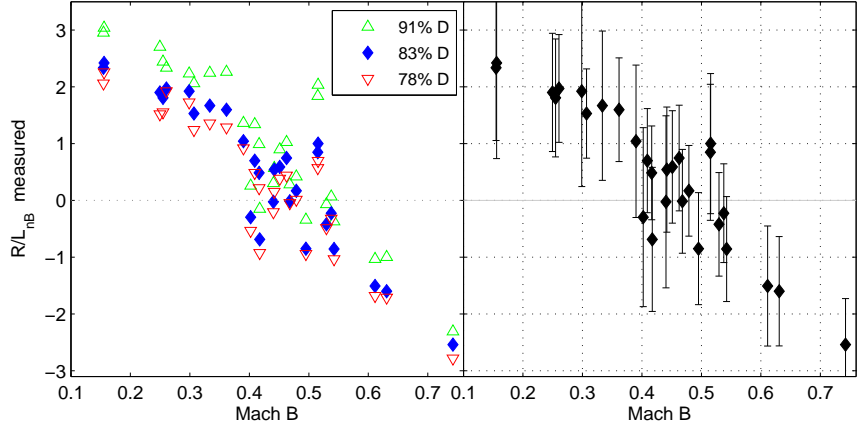}
\caption{Estimated uncertainties in measured $R/L_{nB}$ at mid radius for the entire database. (Left) Boron density gradient sensitivity via the halo contribution to different deuterium concentrations.  (Right) Estimated error bars on the reference points (83\pc D) combining both systematic and random uncertainty estimates using a distribution of spline fits.}
\label{fig.meas_err}
\end{center}
\end{figure}

To quantify the influence of the deuteron density uncertainty, three points from the database were investigated by calculating the boron density with a range of possible deuterium concentrations.  The deuterium concentration was varied between
 \{91\pc, 83\pc and 78\pc\}, with the latter corresponding to a very dirty plasma for AUG (with $Z_{\rm eff}=2.4$ assuming 4\pc He, 1\pc B, 1\pc C and 0.04\pc W).
The effect of these deuterium concentrations on the calculated boron density gradient is shown for the entire database in Fig. \ref{fig.meas_err}.  The results show that the use of the higher (lower) deuterium concentrations systematically increases (decreases) the measured $R/L_{nB}$ at mid radius, due to the variation of the halo contribution to the CX signal.  We also investigated different inversions for the $n_e$ profile and the effect of non-flat $Z_{\rm eff}$ profile; the results do not differ significantly from the set of three deuterium concentrations presented.

To calculate an uncertainty on the logarithmic gradient $R/L_{nB} = -(R / n_B) (\partial n_B / \partial r)$ we separate the global systematic and local random errors for appropriate uncertainty propagation into the gradient quantities.  Since each measurement is a steady state phase of between 0.25 and 1.0 seconds, the fitting error estimate $\varepsilon_{\rm fit}$ is the standard error $(\sim 1 /\sqrt{n_t})$ of the CX fitting errors for each time point (where $n_t$ between 3 and 15 is the number of measurements in the time window).  The standard deviation of the measurements in the time window is considered to be a fluctuation uncertainty $\varepsilon_{\rm fl}$ and we furthermore define a per radial point calibration error $\varepsilon_{\rm ppce}$ which is the distance from the initial spline fit (up to a maximum of 10\pc of the density value).  These per point errors represent the changes from the initial smooth global calibration (in which all lines of sight give smooth profiles), which can occur in different amounts as individual fibres are connected and reconnected subsequent to the calibration.  For each radial point, these uncertainties are combined as
\begin{equation}
  \label{eq.errs}
  \varepsilon_{\rm rp} = \sqrt{\varepsilon_{\rm fit}^2 + \varepsilon_{\rm ppce}^2 + \varepsilon_{\rm fl}^2}.
\end{equation}
Using these radial point errors to add Gaussian noise to each radial point, 10,000 cubic spline fits are generated for every point in the database.  The distribution of $R/L_{nB}$ from the spline fits is itself Gaussian, with mean and median identical to the spline fits on the raw data.  The standard deviation of the logarithmic gradient of the fits $\varepsilon_{\rm rln}$ is combined with the sensitivity to the deuteron profile $\varepsilon_{n_D}$ to produce the error bars $\varepsilon_{\rm tot} = \sqrt{\varepsilon_{\rm rln}^2 + \varepsilon_{n_D}^2}$ shown in Fig. \ref{fig.meas_err} (global systematic uncertainties, e.g. in global calibration or atomic data do not contribute to the logarithmic gradient).

The size of the gradient uncertainties is influenced by the tension on the spline fits.  The value used was chosen (prior to the generation of the randomised fits) as the value which best preserved the trend of the data while removing the influence of the features of individual points.   While the reader may note that there is some subjectivity to this choice, the above exercise demonstrates that fitted logarithmic gradients can be much more uncertain than the absolute density at a given location.  However, the advantage of this dimensionless quantity is that it can be compared with quasi-linear simulations at a single radius.  This allows a quantitative comparison to be made without a calculation of the nonlinear saturation of stiff turbulent fluxes, which are extremely sensitive to the input gradients. 

\section{References}

\bibliographystyle{unsrt} 
\bibliography{aug_boron}

\end{document}